\documentclass[aps,prl,preprint,groupedaddress]{revtex4-1}
\usepackage{epsfig}
\usepackage{graphicx}
\usepackage[caption=false]{subfig}
\usepackage{bm}
\usepackage{color}
\usepackage{float}
\usepackage{mathtools}
\usepackage{amsmath}
\usepackage{comment}

\begin{document}


\title{The AB transition in Superfluid $^3$He}

\author{W. P. Halperin}
\email[]{w-halperin@northwestern.edu}
\author{Man Nguyen}
\author{J.W. Scott}
\author{J.A.Sauls}
\affiliation{Department of Physics and Astronomy, Northwestern University, Evanston IL 60208, USA}

\date{\today}
\begin{abstract}
The discovery of superfluidity in $^3$He in 1971, published in 1972,~\cite{Osh.72,Osh.72b} has influenced a wide range of investigations that extend well beyond fermionic superfluids, including electronic quantum materials, ultra-cold gases and degenerate neutron matter.  Observation of thermodynamic transitions from  the  $^3$He Fermi liquid  to two other liquid phases, A and B-phases,  along the melting curve of liquid and solid $^3$He, discovered by Osheroff, Richardson, and Lee,  were the very first indications of $^3$He superfluidity leading to their Nobel prize in 1996.  This is a brief retrospective specifically focused on the AB transition. 
\end{abstract}


\maketitle

This article  on superfluid $^3$He is a tribute to those early leaders who formed the Low Temperature Laboratory at Cornell University, David Lee and John Reppy, for which this special issue of the Journal of Low Temperature Physics is dedicated, commemorating their 90th birthdays.  

\begin{figure*}[b]
	\hspace*{-0.5cm}
	\includegraphics[scale = 0.2
	]{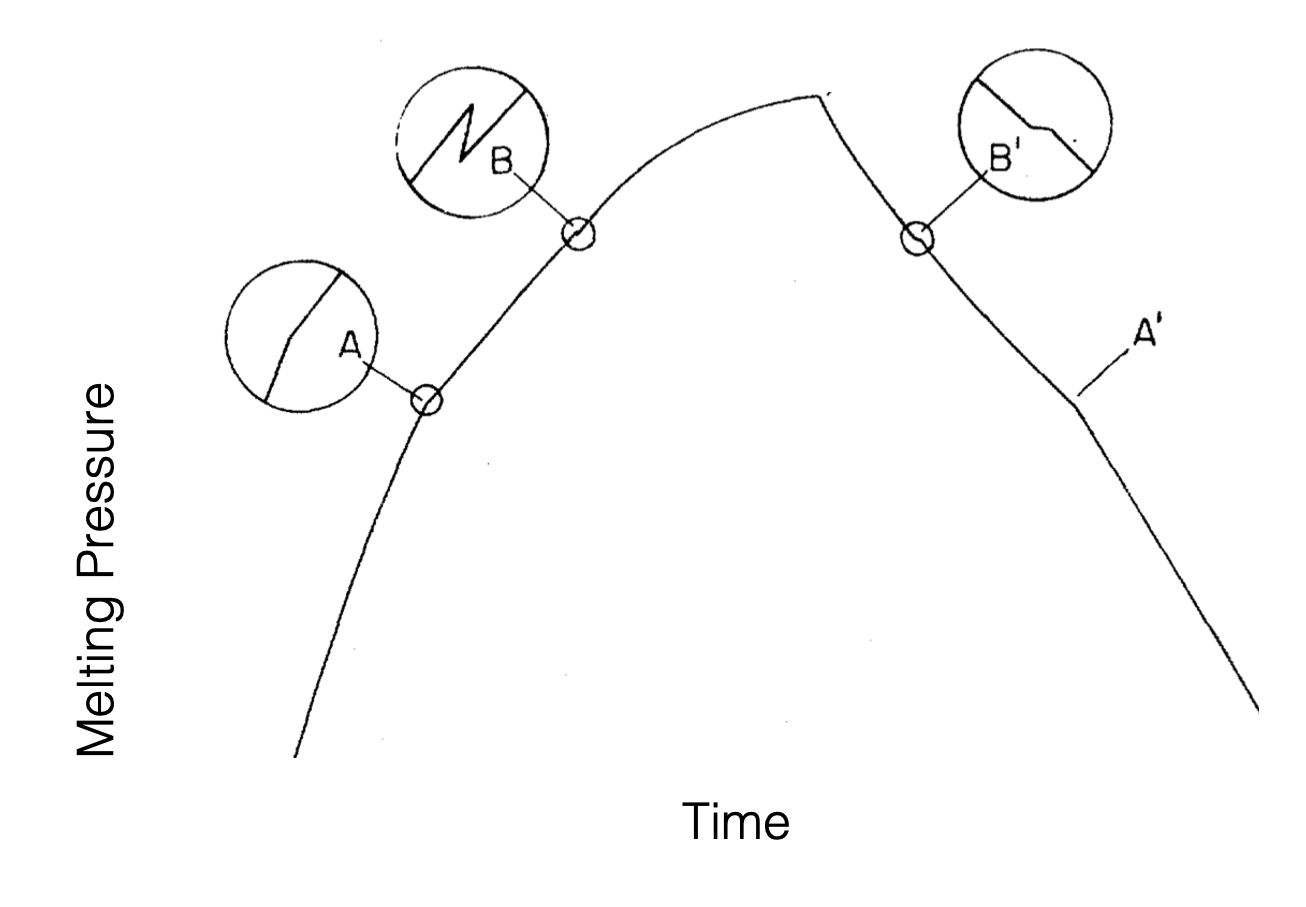}
	\caption{\label{Fig.1} Measurement of melting pressure in a Pomeranchuk refrigerator.  First observation of superfluid  $^3$He by Osheroff , Richardson, and Lee, adapted from Reference~\cite{Osh.72b}.}
\end{figure*}

The world-wide search for quantum phases in liquid and solid $^3$He in the 1960's drove experimental techniques for refrigeration, namely dilution refrigerators and demagnetization cooling of paramagnetic salts, to achieve lower temperature.  A promising alternative method was the pressurization of a mixture, of liquid and solid $^3$He constrained to its melting curve.  It was first noted by Pomeranchuk~\cite{Pom.50} that the coexistence  of liquid and solid, expressed as a pressure-temperature relation, had a negative slope according to the Clausius-Clapeyron equation,

\begin{equation}
dP/dT = \frac{s_l -s_s}{v_l-v_s} 
\end{equation}

\noindent
since the entropy of the solid from nuclear spin disorder, $s_s$, is larger than that of the degenerate Fermi liquid, $s_l$, with corresponding molar volumes $v_s < v_l$. Therefore  pressurization of the mixture, converting liquid to solid, should result in cooling, hence the name Pomeranchuk effect. The Cornell University low temperature laboratory of David Lee and John Reppy, soon joined by Robert Richardson, pursued this technology.  Several versions of refrigerators of this kind were developed and are reviewed by Richardson~\cite{Ric.97}.  All such methods require mechanical devices such as metal bellows, tubes or diaphragms and the limitations from corresponding mechanical heating were unknown.  Despite challenges in thermometry, Pomeranchuk refrigeration demonstrated sufficient cooling power to be effective.  One aspect of thermometry was unambiguous; the higher the pressure the lower the temperature.  The University of Florida group, led by Dwight Adams, invented a high-precision low temperature pressure gauge~\cite{Str.69} and this technology was harnessed by the Cornell group to measure  progress in achieving lower temperatures.  Although qualitative at the time, the $^3$He temperature could be inferred from  NMR of copper or platinum wires, taking advantage of the fact that the magnitude of the NMR signal is an inverse function of absolute temperature.

Late in the fall of 1971,  Osheroff, Richardson, and Lee noted an abrupt but very reproducible decrease in the cooling rate of  $^3$He on the melting curve that they labeled as the A-transition followed by an abrupt drop in pressure at the B-transition, Fig. 1~\cite{Osh.72}.  Osheroff  has provided a detailed personal account of this history and its eventual interpretation marking the discovery of superfluid  $^3$He~\cite{Osh.97}.  The higher temperature A-feature is a sharp decrease in cooling rate that  corresponds to a jump in the heat capacity of the liquid.  At B there is a first order transition from the supercooled  A-phase to the B-phase releasing a pulse of latent heat, hence a drop in pressure. Most importantly, the nuclear magnetic resonance $^3$He frequency shifts~\cite{Osh.72b} in the A-phase were identified by  Leggett to be the signature of liquid $^3$He in an odd-parity superfluid state~\cite{Leg.73}; see the review by Lee and Leggett~\cite{Lee.11}.  With this discovery the vast field of unconventional fermionic pairing was born; a remarkable manifestation of the theory of BCS superconductivity~\cite{Bar.57}, with  appropriate modifications for $^3$He~\cite{And.61,Bal.63,Vdo.63}.  In this article, we emphasize several  important aspects of the discovery of the AB transition; these  are key, not only to the discovery itself, but they still offer further insights  50 years later. 

\begin{figure*}[t]
	\hspace*{-0.5cm}
	\includegraphics[scale = 0.25
	]{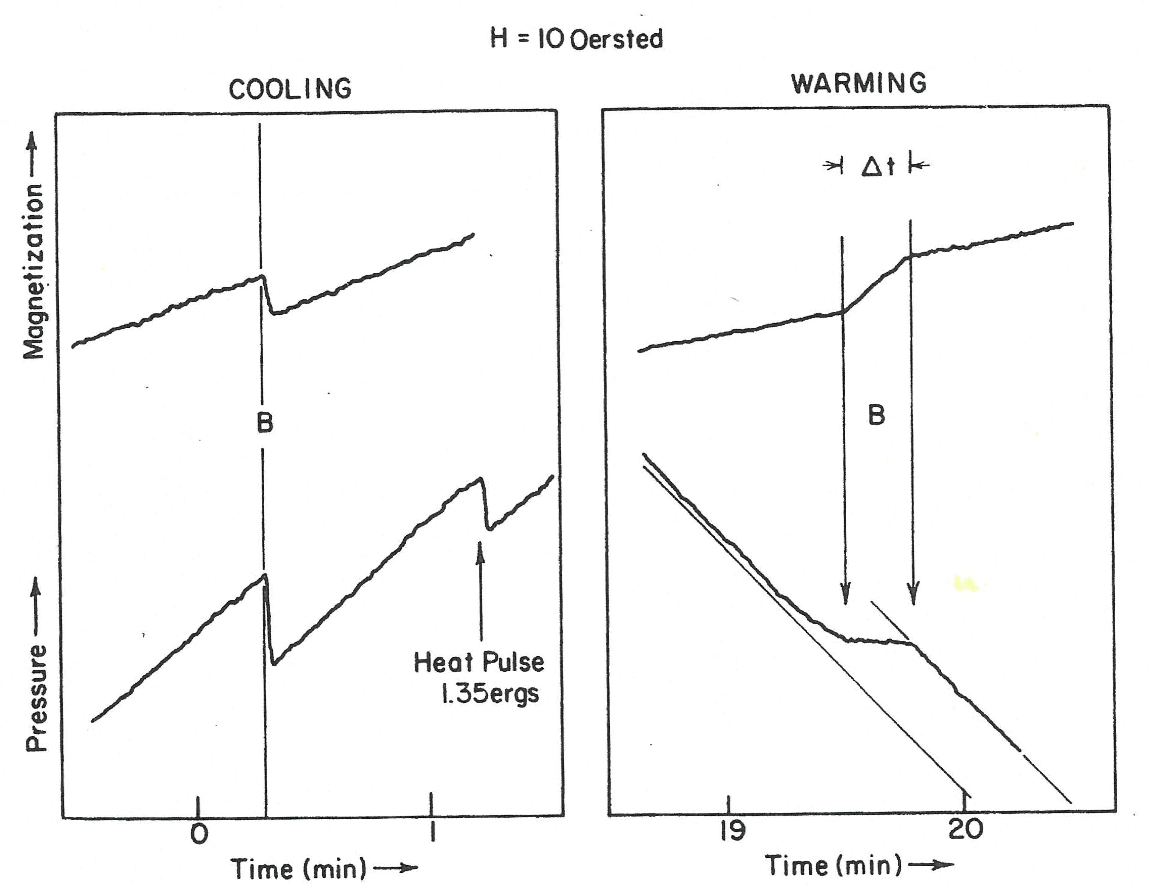}
	\caption{\label{Fig.2} Magnetization and pressure of $^3$He in a Pomeranchuk refrigerator by Lee and collaborators, adapted from Reference~\cite{Hal.73}. Magnetization measurements were performed with a SQUID showing a simultaneous discontinuity with pressure, both cooling and warming through A$\rightarrow$B and B$\rightarrow$A transitions, respectively,  consistent with their first order character.  Importantly, the static magnetization change was found to be consistent with NMR measurements~\cite{Osh.72b}.}
\end{figure*}

\begin{figure*}[b]
	\hspace*{-0.5cm}
	\includegraphics[scale = 0.25
	]{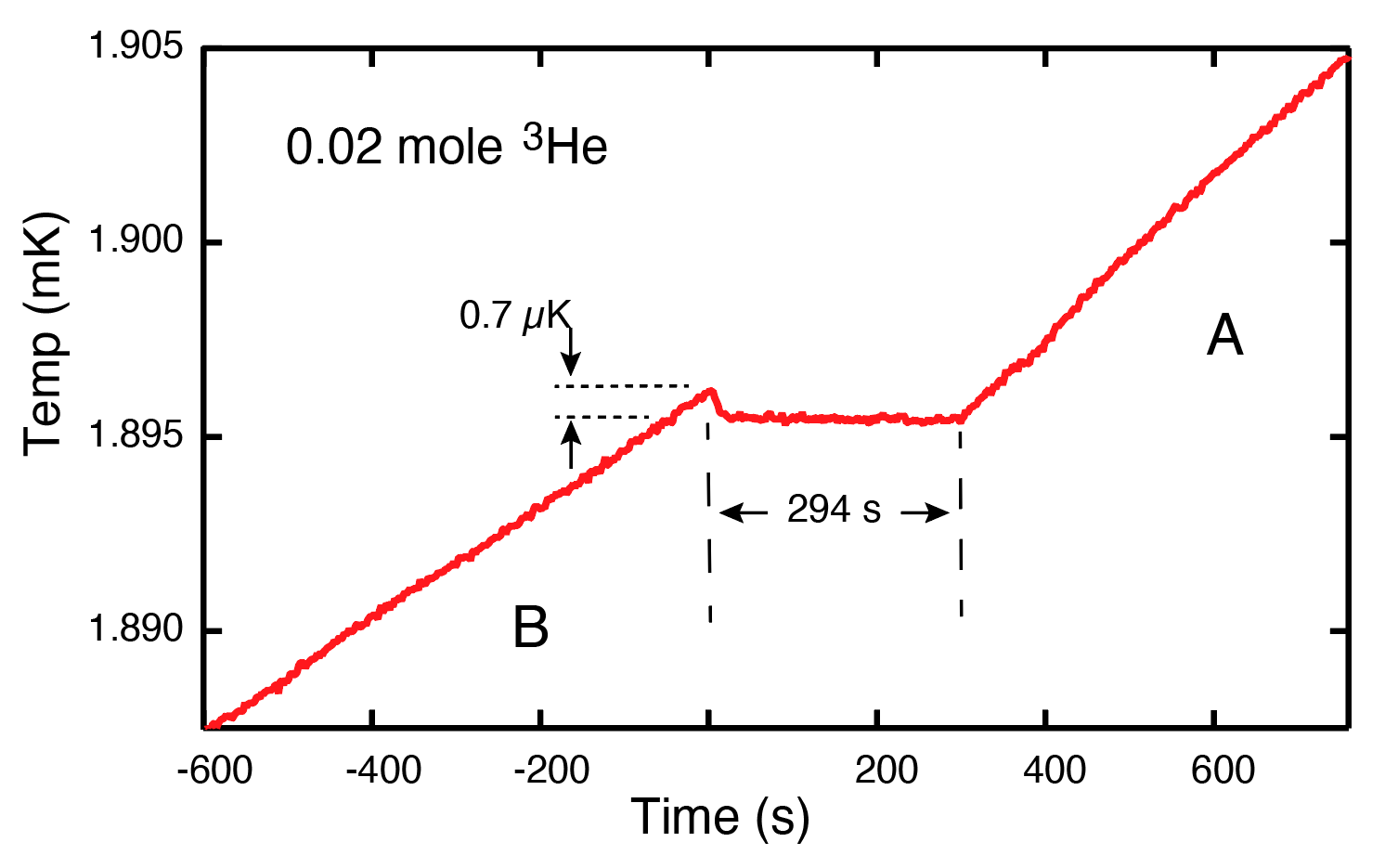}
	\caption{\label{Fig.3} Recent measurement of the B$\rightarrow$A transition in a $^3$He melting curve thermometer based on the PLTS-2000 temperature scale.  The thermometer was cooled below 0.45 mK using nuclear demagnetization and allowed to warm slowly in zero applied field.  Using the latent heat  of 15.4 ergs/mol~\cite{Hal.78}, the heat leak into the liquid (including that from the solid) is 100 pW.  At Northwestern we have found this to be a useful way to check  refrigerator performance~\cite{Ngu.20}. Although very unusual, superheating is responsible for the slight negative temperature jump of $0.7\,\mu$\,K.
	}

\end{figure*}

At that time the predominant goal at Cornell, as well as at the University of Florida (Dwight Adams), University of California at La Jolla (John Wheatley and John Goodkind), and Helsinki University of Technology (Olli Lounasmaa),  was to search for magnetic order in the solid phase where presumably the nuclear spin entropy of the solid must eventually decrease at sufficiently low temperature as required by the third law of thermodynamics.  It was David Lee's inspiration that the search for the ground state behavior of the nuclear spin system of solid $^3$He should be pursued in a mixture of liquid and solid using the Pomeranchuk effect. Nuclear magnetic ordering had been predicted from high temperature data to be in the vicinity of ~2 mK and pursuit of this goal with improvements in refrigeration was central to the work in a number of low temperature laboratories including that of Lee and Richardson.    By comparison, predictions for the superfluid transition temperature ranged widely by six orders of magnitude from $\sim$100\,mK to $10^{-4}$\,mK, as reviewed by Bookbinder~\cite{Boo.71} in 1971.  Without more clarity its search was considered to be an unrealistic goal, but serendipity allowed that the superfluid transition would occur in the same temperature range as the search for the nuclear magnetic ordering laying the foundation for the discovery of superfluid $^{3}$He and a subsequent explosion of research. Initially, the  sharp changes in the time dependence of the melting pressure trace observed by Osheroff $et\,al.$~\cite{Osh.72}, Fig. 1, were interpreted as a manifestation of nuclear magnetic order in the solid. However, one of the doubters of that interpretation was  Goodkind at the University of California San Diego, proposing the possibility of a transition  to superfluidity in the liquid phase~\cite{Lee.11}.  Another was Vvedenski at the Institute of Physical Problems, Moscow~\cite{Vve.72}, who proposed a thermodynamic explanation associated with superfluidity, as noted above, to a decrease in cooling rate attributed to a heat capacity jump in the liquid phase.  Later, Halperin $et\,al.$~\cite{Hal.74,Hal.78} developed a thermodynamic analysis consistent with Vvedenski's idea, but with the central presumption being  that the entire liquid phase of $^3$He on the melting curve is in  thermal equilibrium  with an interface to the solid phase, thereby responding to the well-defined pressure temperature relation of the melting curve with a very short time scale on the order of seconds.  This is in stark contrast to the very long equilibrium times in the solid found to be three orders of magnitude greater than that of the liquid near the superfluid transition~\cite{Hal.78}.

\begin{figure*}[h]
	\hspace*{-0.5cm}
	\includegraphics[scale = 0.2
	]{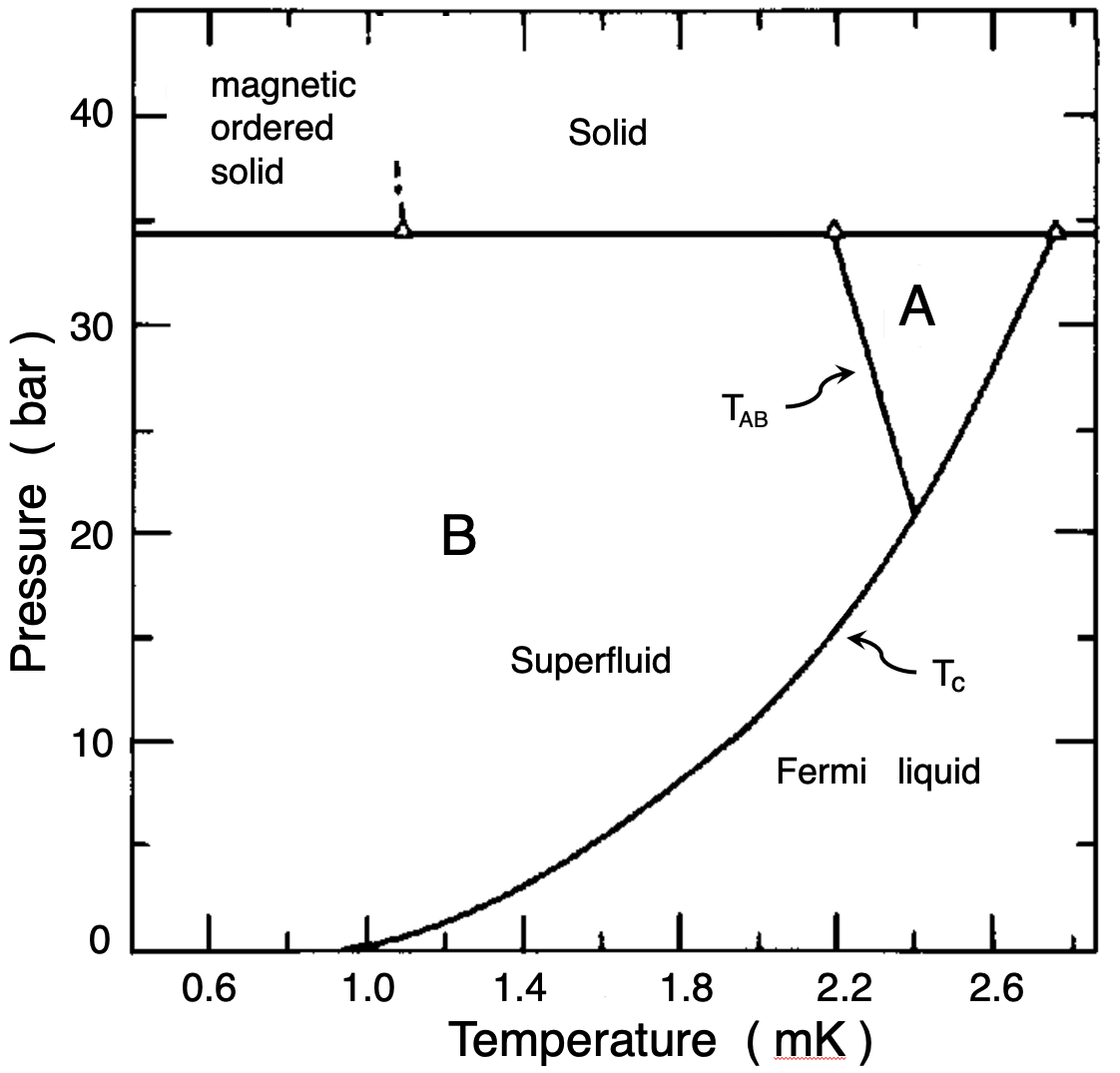}
	\caption{\label{Fig.4} The superfluid $^3$He phase diagram vintage 1978, adapted from Ref.~\cite{Hal.78}. The temperature scale shown here has since been updated by PLTS-2000.	}

\end{figure*}

That thermodynamic model was quantitatively affirmed through the comparison of measurements of the specific heat of the liquid and the latent heat of the B$\rightarrow$A transition~\cite{Hal.74,Hal.78} on the melting curve, Fig. 2,  in comparison with the pure liquid reported by Greywall~\cite{Gre.86}.  If the early experiments on the melting curve using a Pomeranchuk refrigerator had been performed sufficiently slowly with  liquid and solid together in thermal equilibrium, then the superfluid transition would not have been evident.  In fact the development of the $^3$He melting curve thermometer (MCT)~\cite{Scr.70, Gre.82} hinges on the existence of this  well-defined thermodynamic system. For example, Fig. 3 shows the warm-up trace of a small amount of $^3$He, $\sim 0.5$ cm$^3$, in an MCT, by Nguyen $et\, al.$~\cite{Ngu.20} warming through the B$\rightarrow$A transition.  With high resolution in their measurement of temperature they also found evidence for a slight amount of superheating, which is generally thought not to take place at this first order  transition in zero magnetic field. Superheating is discussed by Wheatley~\cite{Whe.75} (section XI) and Leggett and Yip~\cite{Leg.90} (section 3).

On the other hand supercooling is ubiquitous, and is especially strong at high pressure near the melting curve.  A rather general requirement for nucleation is that a seed of the more stable phase in a local volume of the size of the coherence length forms and then expands beyond a critical radius determined by surface energy between the phases and their difference in free energy.  Leggett  argued that B-phase homogeneous nucleation attributed to thermal fluctuations was highly improbable by many orders of magnitude, and that nucleation must be inhomogeneous, most likely initiated from the byproduct of capture of some unknown particle, a process that creates a quantum analogue of the confection called ``baked Alaska".   In this model, the heat released locally from the collision is then radiated by normal quasiparticles, evaporatively cooling an affected volume by their escape from within, creating a temperature inversion similar to a baked Alaska.  Since there is a non-zero probability that one of these events can create a B-phase  in a volume larger than critical, its expansion results in the destruction of the metastable A-phase.  And yet, the background source responsible for this nucleation remains an open problem with speculation including radioactive decay of tritium, cosmic rays, or some form of dark matter; see the review by Schiffer and Osheroff~\cite{Sch.95}.
In fact, exposure to an external neutron flux~\cite{Sch.95,Bau.96,Ruu.96}  has been shown to be an effective means for inducing nucleation and in one case mechanical shock~\cite{Bart.00}.  Furthermore, neutron capture is responsible for vortex formation suggesting that vortices, or other topological defects, might play a role~\cite{Bau.96,Ruu.96}. These, and other investigations of the A$\rightarrow$B transition, have carefully examined the lifetime of the metastable state, its temperature and magnetic field dependence. Recent measurements of supercooling by Parpia at Cornell, but in this case in the near vicinity of the tricritical point (PCP in Fig. 5),  also have interesting unexplained features~\cite{Lot.21}.

\begin{figure*}[t]
	\hspace*{-0.5cm}
	\includegraphics[scale = 0.2
	]{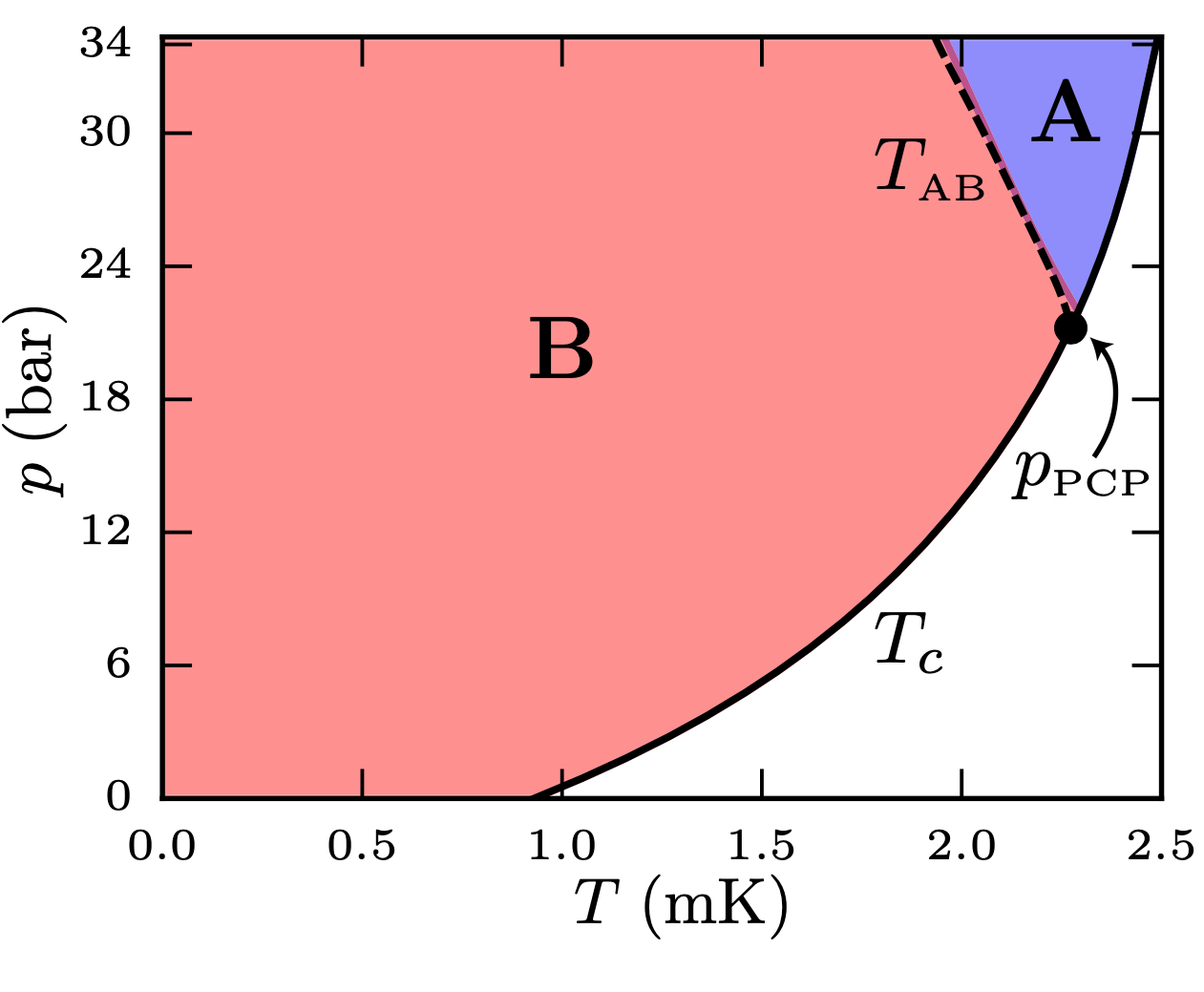}
	\caption{\label{Fig.5} Predicted and experimental temperature dependence of the AB transition pressure, adapted from Reference \cite{Wim.15}.}

\end{figure*}

 Although a theoretical first principles calculation of the transition temperature of superfluid $^3$He is not yet available,  present theory describes many aspects of the behavior of the superfluid, including the relative stability of the A and B-phases.  Within the framework of a Ginzburg-Landau (GL) model, combined with microscopic calculation of strong coupling contributions to the free energy and experimental measurements, Wiman and Sauls~\cite{Wim.15} have recently shown that the temperature dependent properties of the superfluid can be predicted and are in precise agreement with experiment.  This includes  an explanation of the pressure-temperature phase diagram describing the relative stability of the A and B-phases shown in Fig. 5~\cite{Wim.15}, as well as the temperature dependence of the specific heat in the B-phase over a wide range of temperature down to $T/T_c = 0.3$~\cite{Wim.18b}.

The GL Free energy in zero magnetic field can be expanded in terms of the invariants of the order parameter $A$ neglecting the dipole energy:
\begin{eqnarray} F &=& -\alpha \mathrm{Tr}(AA^{\dagger})+
g_{z}H_{\mu}(AA^{\dagger})_{\mu\nu}H_{\nu} + 
\beta_{1}|\mathrm{Tr}(AA^{T})|^{2} \nonumber \\ && +
\beta_{2}[\mathrm{Tr}(AA^{\dagger})]^{2} 
+\beta_{3}\mathrm{Tr}(AA^{T}(AA^{T})^{*}) \nonumber \\ && +
\beta_{4}\mathrm{Tr}((AA^{\dagger})^{2})
+\beta_{5}\mathrm{Tr}(AA^{\dagger}(AA^{\dagger})^{*}).
\label{GL_free_energy}
\end{eqnarray}

\noindent
Here $A^{\dagger}$ and $A^{T}$ are the Hermitian conjugate and transpose of $A$. The 
structure of the order parameter allows five fourth order invariants for which
the coefficients, $\beta_{i}$,  determine the stable superfluid states.  The expansion  coefficients in
the weak-coupling limit are,
\begin{eqnarray} &\alpha={N(0)\over 3} \left({T\over T_{c}}-1\right),\\
&{\beta_{i}\over\beta_{0}} = (-1, 2, 2, 2, -2),  i =1, ..., 5,\\
&\beta_{0}={7\zeta (3)\over 120 \pi^{2}} {N(0) \over (k_{B}T_{c})^{2}},
\end{eqnarray}
\noindent
where the normal density of states at the Fermi energy is  $N(0)$, $k_{B}$ is the  Boltzmann
constant and
$\zeta(x)$ is the Riemann zeta function. However, $^{3}$He is not a weak
coupling superfluid and strong  coupling effects increase with pressure.  In the absence of strong coupling the superfluid would be the isotropic state, corresponding to the B-phase as predicted by Balian and Werthamer~\cite{Bal.63} and independently by Vdovin~\cite{Vdo.63}. But at high pressure the anisotropic axial state, {\it i.e.} the A-phase, is stabilized by
strong coupling~\cite{Ser.83}.  These corrections have a negligible influence on $\alpha$~\cite{Sau.81}, but they
 significantly modify the $\beta_{i}$'s. Expressions for the combinations of the  $\beta_i$ that can be determined from experiment are reviewed by Choi {\it et al.}~\cite{Cho.07}.  In particular the condition for equilibrium between A and B-phases is $\beta_{A} \equiv \beta_{245} = \beta_{B} \equiv \beta_{12} + \frac{1}{3}\beta_{345}$, where the subscript sequence indicates the sum of corresponding $\beta$'s.  In particular the polycritical point (PCP in Fig. 5) is defined by the $\beta_i(T_c)$.  Calculations of strong coupling corrections have been performed for model potentials by Sauls and Serene~\cite{Sau.81} and in more recent work by Wiman and Sauls~\cite{Wim.15,Wim.18b}.  A major advance regarding prediction of behavior of the superfluid at temperatures below $T_c$ was implemented by them incorporating the temperature dependence of strong coupling   calculated by Serene and Rainer who found these corrections to the free energy to  scale with temperature~\cite{Ser.83}.  This greatly extends the range of validity of the theory correctly accounting for the pressure-temperature dependence of the equilibrium curve $T_{AB}$ in Fig. 5, as well as the temperature dependence of the B-phase specific heat~\cite{Wim.15,Wim.18b}.

In summary, among many advances in research on superfluid $^3$He which continue to the present, the study of the A$\rightarrow$B transition is particularly fascinating and for which there are important open problems.  We are grateful for support from the National Science Foundation (WPH, Grant No. DMR-1903053 and JAS, Grant No. DMR-1508730). One of us (WPH) acknowledges being engaged in $^3$He superfluid physics, Fig. 2, with co-author David Lee in the first measurements of the static magnetization and latent heat at this transition~\cite{Hal.73}. 

%

\end{document}